\documentclass[11pt,english,sort,super,compress]{article}
\usepackage{multicol}
\usepackage{wrapfig}
\usepackage{dcolumn}
\usepackage{bm}

\usepackage{booktabs}
\usepackage{lineno}
\usepackage{hyperref}

\usepackage{graphicx}
\usepackage{amssymb,amsmath}
\usepackage{bbm}
\usepackage{float}
\usepackage{bm}
\usepackage{setspace}
\usepackage{parskip}

\usepackage{algorithm}  
\usepackage{algorithmic} 
\usepackage{ulem} 

\usepackage{xcolor}
\definecolor{midnightblue}{cmyk}{1,1,0,0.1}
\definecolor{forestgreen}{cmyk}{0.76,0,0.26,0.5}

\usepackage[a4paper]{geometry}
\usepackage{epsfig}
\usepackage[numbers]{natbib}
\usepackage{lineno}
\usepackage{hyperref}
\hypersetup{
    bookmarks=true,         
    unicode=false,          
    pdftoolbar=true,        
    pdfmenubar=true,        
    pdffitwindow=false,     
    pdfstartview={FitH},    
    pdftitle={My title},    
    pdfauthor={Author},     
    pdfsubject={Subject},   
    pdfcreator={Creator},   
    pdfproducer={Producer}, 
    pdfkeywords={keyword1} {key2} {key3}, 
    pdfnewwindow=true,      
    colorlinks=true,       
    linkcolor=midnightblue,          
    citecolor=magenta,        
    filecolor=midnightblue,      
    urlcolor=midnightblue,          
}

\usepackage{listings}
\definecolor{mygreen}{rgb}{0,0.6,0}
\definecolor{mygray}{rgb}{0.5,0.5,0.5}
\definecolor{mymauve}{rgb}{0.58,0,0.82}

\makeatletter
\usepackage{lineno}
\makeatother
\usepackage{babel}

\usepackage{mathabx}

\newcommand{\be}{\begin{equation}}
\newcommand{\ee}{\end{equation}}

\newcommand{\bk}{{\bm{k}}}

\newcommand{\bq}{{\bm{q}}}

\newcommand{\ii}{{\mathrm{i}}}
\newcommand{\tr}{{\text{tr}}}

\begin{document}

\par

\begin{spacing}{1.618}
\begin{center}{\LARGE{A non-iterative method for the vertex corrections of the Kubo formula for electric conductivity}}

\vspace{0.2 in}
Yi-Wen Wei$^{1}$, Chao-Kai Li$^1$,  Yuchuang Cao$^1$, Ji Feng$^{1,2,3,*}$ 
\vspace{0.2 in}

$^{1}$\textit{International Center for Quantum Materials, School of Physics,\\
Peking University, Beijing 100871, China }
\\$^{2}$\textit{Collaborative Innovation Center of Quantum Matter, Beijing, P. R. China} \\
$^{3}$\textit{CAS Center for Excellence in Topological Quantum Computation, University of Chinese Academy of Sciences, Beijing 100190, China}\\
$^{*}$jfeng11@pku.edu.cn

\end{center}

\bigskip

\begin{abstract}
In computing electric conductivity based on the Kubo formula, the vertex corrections describe such effects as anisotropic scattering and quantum interference and are important to quantum transport properties. These vertex corrections are obtained by solving Bethe-Salpeter equations, which can become numerically intractable when a large number of $\bk$-points and multiple bands are involved. We introduce a non-iterative approach to the vertex correction  based on rank factorization of the impurity vertices, which significantly alleviate the computational burden. We demonstrate that this method can be implemented along with effective Hamiltonians extracted from  electronic structure calculations on perfect crystals, thereby enabling quantitative analysis of quantum effects in electron conduction for real materials.
\end{abstract}

\bigskip
\bigskip


\section{Introduction}
The Kubo formula establishes a fundamental link between near-equilibrium processes and equilibrium correlation functions, enabling a microscopic description of various transport phenomena~\cite{Kubo57}.  When applied to electronic transport in the presence of external fields, the Kubo formula for current response leads to the quantum mechanical formulation of electrical conductivity~\cite{Kubo57,Green54}.  the single most important property of materials. Indeed, electron conduction by itself displays an extraordinarily wide range of unusual phenomena, which sometimes reveal highly nontrivial physical principles. Of particular interest to quantum transport phenomena is the localization of electrons in the presence of  disordered impurities in an otherwise ideal metallic system, where electrons tend to be more localized compared to classical processes. Localization of electrons is crucial to the discussions of a variety of problems, including weak localization, magnetoresistance, quantum Hall effect and topological transport properties. Such electronic localization is in essence a consequence of quantum interference, which in the Kubo formula is embodied in a set of vertex corrections, including those from the maximally-crossed diagrams~\cite{Langer66,Anderson79,Gorkov79,Altshuler80}.

The vertex correction corresponding to a partial sum of the maximally-crossed diagrams is obtained by solving the Bethe-Salpeter equation, which is a 4-point Dyson equation. Successful theories have been developed by considering low-energy effective Hamiltonians and only the leading order contribution to vertex correction, to account for, for example, localization or the absence of it in novel Dirac and Weyl semimetals~\cite{Suzuura02,Lu14,Lu15}.  It is, however, more of a challenge to numerically solve the full vertex correction for a generic band structure of a realistic material, where it becomes unlikely to achieve simplification (based on symmetry and otherwise). This challenge arises plainly from the fact that when a large number of $\bk$-points and multiple bands are involved the vertex is a non-sparse matrix of an enormous size, which practically defies obvious methods for solving the matrix equation. This is so, despite the fact that high-quality electronic structure of real materials can routinely be obtained based on state-of-the-art electronic structure methods. Especially, Wannier functions~\cite{Marzari12} have become a standard technique for {down-folding} the band structure and Brillouin zone interpolation, allowing us to compute a wide range of physical properties conveniently~\cite{Mostofi09, Noffsinger10, Shelley11, Pizzi14, Assmann16}.

In this paper, we present a method to compute the vertex corrections arising in the calculation of electric conductivity for a crystalline system in a non-iterative approach, which can greatly alleviate the computational burden and boost efficiency. The paper is organized as follows. We will begin with a brief recapitulation of the Kubo's theory for electrical conductivity, leading to the vertex corrections, whereupon the essential formulae and notations are introduced. An analysis of the vertex in the presence of disordered, uncorrelated impurities in a fairly generic form will be presented. It is revealed that the impurity vertex can have fewer degrees of freedom than its apparent dimension, owing to the short-rangedness of impurity potential. This crucial fact leads  to an algorithm based on rank factorization, which reduces the dimension of a Bethe-Salpeter equation.  The computational load of the rank factorization is further reduced by employing a projective singular value decomposition method.  The stability and efficiency of our method for the vertex correction are examined, and compared with such direct methods as matrix inversion and biconjugate gradient method. We show that the computational complexity of our method is formally $\sim O(n^2)$ and indeed scales well in our implementation, whereas that of matrix inversion or biconjugate gradient method is ${\sim O(n^{3})}$. Finally, we present two concrete examples, applying this method to 2-dimensional monolayer lead (Pb) with a hexagonal lattice and 3-dimensional face-centered cubic (fcc) Pb  to obtain the vertex corrections to conductivity. These illustrative examples show that our method can be efficiently applied to obtain the aforementioned vertex corrections to conductivity, thereby quantitatively analyzing weak localization for real materials. 

\section{Vertex corrections in conductivity}
In this section,  the problem that we intend to solve is introduced, along with essential notations. The electronic Hamiltonian and the nature of the impurity potential will be clarified. Starting with the Kubo formula for electric conductivity, the vertex corrections are introduced, which in the diagrammatic expansion of the disorder average lead to the Bethe-Salpeter equations. In this section, the vertex corrections in electric conductivty are introduced diagrammatically to highlight the basic structure of the theory, and their algebraic forms are postponed to the next section. Also included is a nonexhaustive discussion of available methods for computing electric conductivity and the vertex corrections. 

We will work primarily in the Bloch representation, and operators are expanded in the eigenstates of the self-consistent mean-field single-particle electronic Hamiltonian, $H$, for instance from various implementations of the density-functional theory (DFT). Thus, the Hamiltonian is only effectively non-interacting, as part of the interactions and correlation are already captured by the mean-field theory. Within the static-lattice approximation, the Bloch theorem applies, and we shall assume the knowledge of the solution to the Schr\"odinger equation
\be
H\psi_{n\bk} = \varepsilon_{n\bk}\psi_{n\bk},
\label{eq:Bloch}
\ee
where $\psi_{n\bk}$ is the Bloch eigenstate for the $n$th band at quasimomentum $\hbar\bk$. From the Bloch functions, Wannier functions can be obtained to afford a lattice representation in which the impurity potential can be conveniently written. Associating the Bloch function with the Fermion operator $c_{n\bk}$, the Fermion operator for destroying an electron in orbital $a$ at lattice point $\bm{R}$ is 
\be
c_{a\bm{R}} = \frac{1}{\sqrt{n_k}}\sum_{n\bm k} e^{-\mathrm{i} \bm k\cdot\bm{R}} U_{an} (\bk)c_{n\bm k},
\ee
where $U(\bk)$ is a unitary matrix obtained from various approaches to the constructions of Wannier-type functions~\cite{Marzari12}. Here, $n_k$ stands for the number of $\bm k$ points, which is equivalent to the number of lattice sites under the Born-von K\'{a}rm\'{a}n boundary condition.

We shall confine ourselves to the problem of  electric conductivity in the presence of randomly distributed elastic scatterers. The general form of one-body impurity potential written for an Wannier-type basis on a lattice is
\be
H' =\sum_{a\bm{R}}\sum_{b\bm{R'}}\sum_{\bm{R}_i} V_{ab}(\bm{R}-\bm{R}_i,\bm{R}'-\bm{R}_i) c_{a\bm{R}}^\dagger c_{b\bm{R}'} + H.c.,
\label{eq:imp}
\ee
where $\bm{R}(\bm{R}')$ is the lattice vector, whereas $a,b$ label the Wannier-type orbitals. $\bm R_i$ is a lattice site an impurity resides. This form of impurity potential admits of hopping between orbitals on the same and different lattice sites. 
We will focus on short-ranged impurities, which have been widely employed to model scattering in electronic transport, especially for the vertex corrections to conductivity~\cite{Ando2002,Ando1998,Lu14,Lu15}.
It is assumed that the impurity potential has a cutoff interaction range $R_c$.
i.e. $V_{ab}(\bm{R}-\bm{R_i},\bm{R}'-\bm{R_i})=0$, if $|\bm{R}-\bm{R_i}|>R_c$ or $|\bm{R}'-\bm{R_i}|>R_c$. Then in the Bloch representation, the general impurity potential have the following matrix form
\be
H'_{\bk\bk'} =
S_{\bk-\bk'} U^\dagger_\bk V_{\bk\bk'}U_{\bk'}
\equiv S_{\bk-\bk'} W_{\bk\bk'}
\ee
where $S_\bk=n_k^{-1}\sum_{\;i}e^{\ii \bk\cdot\bm{R}_i}$, and $V_{a\bk,b\bk'}=\sum_{\bm{R},\bm{R'}}V_{ab}(\bm{R},\bm{R}')e^{\ii (\bk\cdot\bm{R}-\bk'\cdot \bm{R}')}$. 

We now introduce the Kubo formula for the electric conductivity in the presence of disordered impurities. The linear response approach is naturally couched in the machinery of Green's functions, owing to its perturbative nature. An advantage of this method is that for impurity scattering its result contains quantum corrections to conductivity that are responsible for weak localization effect, which is not included in the usual Boltzmann conductivity. The longitudinal electrical conductivity in the long-wavelength limit derived from the general Kubo formula for a non-interacting electronic system in the presence of an ensemble of disordered scatterers is ~\cite{Kubo57,Greenwood1958,Rammer98,Economou90,Suzuura02}
\begin{eqnarray}
\sigma^{xx} (\omega)=
\frac{e^2}{\Omega}\int \frac{d\varepsilon}{2\pi} \frac{f(\varepsilon)-f(\varepsilon+\hbar\omega)}{\omega} \times\tr \langle v^x G^R(\varepsilon) v^x G^A(\varepsilon+\hbar\omega)\rangle_{\text{d}}, 
\end{eqnarray}
where $\omega$ stands for frequency, $\bm{v}$  velocity operator,  $f(\varepsilon)$ the Fermi-Dirac distribution, $\Omega$ the volume of the system. 
$G^{R/A}(\varepsilon  ) = ( \varepsilon - H-\Sigma^{R/A}(\varepsilon))^{-1}$ are the retarded/advanced single-particle Green's functions evaluated in the presence of a given disorder configuration, which results in corresponding self energy $\Sigma^{R/A}(\varepsilon)$. And $\langle...\rangle_{\text d}$ indicates disorder averaging. Upon averaging over an ensemble of scatterer configurations, the current response is expanded in terms of scattering vertices into an infinite series of distinct current bubbles. The bare (zeroth order) current bubble yields the Drude conductivity~\cite{Rammer98,Economou90}, whereas other current bubbles contain vertex corrections arising from impurity induced scattering between electron and hole excitations~\cite{ward}.
\begin{figure}
	\includegraphics[width=80 mm]{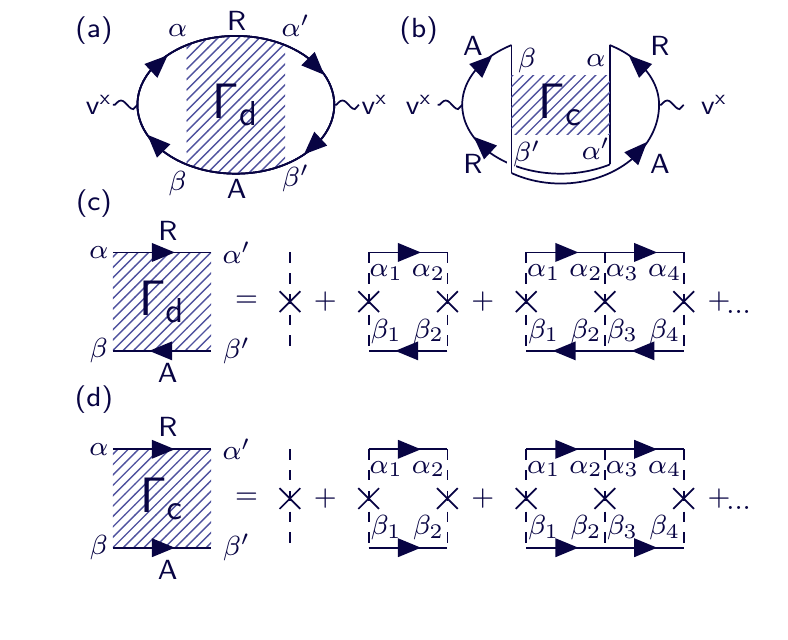}\centering
	\caption{\label{fig:001}Current bubbles with the vertex correction from (a) the Ladder diagram, and (b) the maximally-crossed diagram. The wiggly lines are bare velocity vertices. $\alpha,\alpha_i,\beta,\beta_i$ are indices which combine band and momentum. The arrowed solid lines are disorder-averaged retarded or advanced single-particle propagators, as labeled by R and A. An impurity vertex is represented by a dashed line with a cross in the center. The vertices shown as shade blocks are, respectively,  the (a) diffuson and (b) cooperon. And the corresponding Bethe-Salpeter equations are depicted diagrammatically in (c) and (d). }
\end{figure}
In weak-scattering regime, the leading corrections to bare current bubble arise from simultaneous scattering of propagating quasiparticles by the same impurity site. Neglecting multiple scattering, two classes of diagrams are usually calculated: ladder diagram and maximally-crossed diagram as shown in Fig.~\ref{fig:001}(a) and (b), respectively. The ladder diagram shown in Fig.~\ref{fig:001}(a) is composed of a pair of bare current vertices connected by 4-point diffuson vertex, $\Gamma_{\text d}$, with countercurrent electron and hole propagators. The diffuson vertex displays diffusive poles, with respect to momentum exchange, from successive elastic scattering by impurities. The partial sum of the ladder diagrams amounts to the usual vertex correction for anisotropic scattering,  which replaces apparent scattering time with transport relaxation time and leads to the equivalent of the Boltzmann conductivity. With the implementation of semiclassical Boltzmann transport formalism, it is already possible to evaluate for real materials this part of the bulk conductivity via  maximally-localized Wannier orbitals obtained from density-functional theory calculations without further vertex corrections~\cite{Pizzi14}. and the ballistic conductance of a finite structure within the Landauer's formulation~\cite{Calzolari04}.

The maximally-crossed diagrams~\cite{Langer66, Gorkov79,Anderson79,Altshuler80} shown in Fig.~\ref{fig:001}(b) are important to the discussion of quantum correction to the conductivity, arising from interference  owing to the wave nature of electrons. The maximally-crossed diagrams can be depicted with an exact diagrammatic twist in a time-reversal invariant system, by reversing one of the single-particle Green's function line. This leads to a pair of current vertices connected to the 4-point cooperon vertex, $\Gamma_c$, a propagator in the particle-particle channel analogous to superconductivity. $\Gamma_c$ is also a ladder diagram but with concurrent single-particle lines, and therefore it displays diffusive poles with respect to the total momentum of scattering. The cooperon diagram is especially relevant to the phenomena of weak localization or weak antilocalization in the presence of disorder when time-reversal symmetry is preserved, which can be suppressed upon the removal of time-reversal symmetry by magnetic impurity or external magnetic field. Weak localization and weak antilocalization have dramatic impact on the transport properties at low temperatures particularly in low-dimensional quantum systems. 

Given our access to accurate electronic structures for real materials from various methods, it is highly desirable to be able to evaluate electrical conductivity including the vertex corrections for real materials, which requires the evaluation of the diffuson vertex $\Gamma_d$ and cooperon vertex $\Gamma_c$. Whereas useful and oftentimes analytical results can be obtained for model Hamiltonians with high symmetry in conjunction with simple scattering potential~\cite{Ando2002,Lu14}, the task is computationally challenging for a generic materials with more complex scattering potentials, for which multiple bands and a large $\bm k$-point set are typically involved in the calculations and limited simplifications, if any, can be made from symmetry.

With this outline of the problem, a non-iterative approach to the Bethe-Salpeter equation for the vertex correction to electric conductivity is proposed and shown to alleviate the burden of computation.  In the subsequent section, an analysis reveals that for scattering potential of finite range the rank of scattering vertex in the Bethe-Salpeter equation can be significantly lower than its dimensions. This fact  will then be systematically exploited to simplify the solution of the Bethe-Salpeter equation.

\section{Analysis and algorithm}
As introduced in the previous section, the diffuson and cooperon vertices can both be expressed as infinite series. The series can be folded into integral equations  referred to as Bethe-Salpeter equation, as  diagrammatically shown in Fig.~\ref{fig:001}(c) and (d). Expanded in a complete set of Bloch functions $\psi_{n\bk}$ as in Eq.~(\ref{eq:Bloch}) with the abbreviations, $\alpha,\beta,\alpha',\beta',\alpha_i, \beta_i \rightarrow n\bk$, the Bethe-Salpeter equations can be written as a matrix equation~\cite{ward},
\begin{eqnarray}
\Gamma_{\alpha\alpha'\beta\beta'}(\varepsilon,\omega)
=\Gamma_{\alpha\alpha'\beta\beta'}^0 + \sum_{\alpha_1\alpha_2\beta_1\beta_2}\Gamma^0_{\alpha\alpha_1\beta\beta_1} G^R_{\alpha_1\alpha_2}(\varepsilon+\hbar\omega)\nonumber
\times G^A_{\beta_1\beta_2}(\varepsilon)\Gamma_{\alpha_2\alpha'\beta_2\beta'}(\varepsilon,\omega),
\label{eq:bse01}
\end{eqnarray}
where the bare impurity vertex $\Gamma^0$ is a matrix composed {of} elements corresponding to simultaneous scattering off of a single impurity by electron and hole, 
\be
\Gamma^0_{\alpha\alpha'\beta\beta'}=\langle H'_{\alpha\alpha'}H'_{\beta\beta'} \rangle_{\text{d}},
\ee
where $H'$ is the static impurity potential as given in Eq. (\ref{eq:imp}). The diffuson and cooperon's Bethe-Salpeter equations differ in quasimomentum conservation, dictated by the momentum flows of the single-particle propagators in the conductivity bubbles, respectively, as
\be
\bk_{\alpha}\mp\bk_{\beta}=\bk_{\alpha'}\mp\bk_{\beta'}\equiv \bq.
\ee
Note that the disorder averaged single-particle propagators, $G_{\alpha\alpha'}^{R/A}(\varepsilon)$, are diagonal in $\bk$, i.e. $\bk_{\alpha}=\bk_{\alpha'}$, but are in general non-diagonal in band indices due to self-energy insertion. Therefore,
the Bethe-Salpeter equation, in consideration of the quasimomentum conservation, is decoupled into separate equations for each $\bq$,

\be 
\Gamma_{mn\bk,m'n'\bk'}=
\Gamma_{mn\bk,m'n'\bk'}^0+
\sum\limits_{m_1n_1\bk_1\atop m_2n_2\bk_2}
\Gamma_{mn\bk,m_1n_1\bk_1}^0
K_{m_1n_1\bk_1,m_2n_2\bk_2}
\Gamma_{m_2n_2\bk_2,m'n'\bk'},
\label{eq:bse}
\ee
where 
$
K_{m_1n_1\bk_1,m_2n_2\bk_2}=\delta_{\bm{q},\bk_1\mp \bk_2} G^R_{m_1m_2}(\bk_1)G^A_{n_1n_2}(\bk_2).
$
Here, $\alpha,\beta$, et cetera have been replaced by $m\bk, n\bk'$, et cetera.
The energy dependencies of the vertices and propagators are omitted for brevity, as they can be unambiguously inferred.

To solve the Bethe-Salpeter equation with a direct method, either by matrix inversion or by an iterative procedure such as biconjugate gradient method~\cite{Vance1984}, poses a significant computational challenge  especially for 3-dimensional systems with complex Fermi surfaces. Usually a large number of $\bm k$-points are necessary to ensure adequate sampling of the Fermi surface, which means the dimension $n$ of the complex-valued non-sparse matrices $\Gamma^0$ and $\Gamma$ is large, where $n= n_b^2n_k$, and $n_b$ and $n_k$ are respectively the numbers of bands and $\bm{k}$-points. This can hinder the use of standard libraries for numerical linear algebra to perform the matrix inversion because of excessive memory load. In an iterative procedure, such as biconjugate gradient method, the equation is solved in a column-wise fashion to avoid cramming the memory. However, both of these direct methods have a nominal computational complexity of $O(n^3)$, i.e., the asymptotic complexity of (na\"ive) matrix multiplication. Even when adopting the Strassen algorithm with a reduced numerical stability~\cite{Miller75}, the complexity exponent is $\log_2 7\approx$ 2.807. Although coherent-potential approximation has been used to simplify the vertex corrections for both equilibrium~\cite{Velicky69} and non-equilibrium~\cite{Ke08} theories the maximally-crossed diagrams are absent naturally in this approximation~\cite{Economou90}. It is highly desirable, therefore, to find an approach to Eq. (\ref{eq:bse}) that is both computationally efficient and memory thrifty.

Our approach is based on the fact that the rank of $\Gamma^0$ for the impurity potential described above is no greater than $n_b^2 \times \min\{n_k, n_r^2\}$,  where $n_r$ is the number of lattice sites within the range $R_c$. The claim is proved as follows. The  disorder averaging of the impurity vertex only reduces the geometric structure factor of the impurities to a constant factor, $\langle S(\bk-\bk')S(\bk'-\bk)\rangle \rightarrow n_i/n_k^2$ where $n_i$ is the number of impurities, for electron-hole scattering by the same impurity site. For a given transfer or total momentum $\bq$, the bare vertex $\Gamma^0$ is composed of $n_k\times n_k$ blocks, which follows from quasimomentum conservation. Each individual block, $\Gamma^0_{\bk\bk'}\in \mathbb{C}\,^{n_b^2\times n_b^2}$, corresponds to a pair of momenta on the $\bk$-mesh used in the computation.  Then for a given $\bq$, we have a single impurity vertex  given by
\be
\Gamma^0_{\bk\bk'}(\bq) = \frac{n_i}{n_k^2}\,
W_{\bk,\bk'} \otimes W_{\bq\pm\bk,\bq\pm\bk'},
\ee
where $\otimes$ stands for tensor multiplication, here with respect to band indices. The full $\Gamma^0$ matrix is a tensor product with respect to $\bk$ of two vectors composed of matrix blocks. 

Now we construct a tensor product $\Lambda$ to whom $\Gamma^0$ is a submatrix. Explicitly, we define a matrix $\Phi \in \mathbb{C}^{n_bn_k\times n_bn_r} $: $\Phi_{m\bk,a\bm{R}}=e^{\ii\bk \cdot \bm{R}}U_{ma}^*(\bk)/\sqrt{n_k}$. The rank of $\Phi$ is no greater than min$\{n_bn_k,n_bn_r\}$. Also note that the rank of  $V$ is no greater than $n_bn_r$, for the rank of a matrix cannot exceed the smaller of its dimensions. Then the rank of $ \Phi V \Phi^{\dagger} $ is no greater than min$\{n_bn_r,n_bn_k\}$, from the rank inequality of matrix product. Define another square matrix $\Lambda$ with dimensions $n_b^2n_k^2$, 
\be
\Lambda =  \frac{n_i}{n_k^2}(\Phi V\Phi^{\dagger})\otimes (\Phi V \Phi^{\dagger}),
\ee
meaning that
$
\Lambda_{ij,kl}= \frac{n_i}{n_k^2}(\Phi V \Phi^\dagger)_{ik}(\Phi V \Phi ^\dagger)_{jl}.
$
From the rank inequality of tensor product, we see that the rank$(\Lambda)\le$ min$\{n_b^2n_k^2,n_b^2n_r^2\}$. It only remains to recognize that $\Gamma^0$ is a momentum-conserving submatrix of $\Lambda$ of order $n_b^2n_k$, since 
\[
\Gamma^0_{mn\bk,m'n'\bk'}(\bq)
=
\Lambda_{m\bk,n\bq-\bk;
	m'\bk',n'\bq-\bk'}.
\] 
It is then implied that rank of $\Gamma^0$ is no greater than rank$(\Lambda)$ for  a submatrix cannot have higher rank. Therefore, let $r\equiv \text{rank} (\Gamma^0)$, and we have
\be\label{eq:r}
r\le \text{min}\{n_b^2n_k, n_b^2n_r^2\}.
\ee

This is first key result of this paper, showing that given a general form of the scattering potential on a lattice model Eq. (\ref{eq:imp}), the independent degrees of freedom  of  the impurity vertex is less than min$\{n_b^2n_k,n_b^2n_r^2\}$. When  $n_r^2 < n_k$, the rank of $\Gamma^0$ is no greater than $ n_b^2n_r^2$, regardless of the number of $\bk$-points included in the calculation. In particular, $r \le n_b^2$, if the impurity scattering is completely localized, i.e., $n_r=1$.

That the non-sparse $\Gamma^0$ matrix has lower rank than its apparent dimensions  if $n_r^2 < n_k$ can be exploited to facilitate the solution of the Bethe-Salpeter  equations. The bare vertex $\Gamma^0$, which is a square matrix with $r$ non-zero eigenvalues $\{\gamma_i|i=1,...,r\}$, is bound to have the compact singular value decomposition
\be
\Gamma^0= P D^0Q^\dagger
\ee
where $D^0 \in \mathbb{C}^{r\times r}$ with $D^0_{ij}=\gamma_i\delta_{ij}$, and $P,Q\in \mathbb{C}^{n\times r}$ have full column rank. Here, ran$(P)$ (i.e., the range of $P$) and ran$(Q)$ are, respectively, $r$-dimensional subspaces  of the $n$-dimensional linear space. The subspace completeness is ensured by the equality of the number of orthogonal vectors and the rank $r$. $P$ and $Q$ are both column orthogonal,
\be
P^\dagger P=I_{r\times r}=Q^\dagger Q.
\ee
It may be noted that $P$ and $Q$ are neither row orthogonal nor necessarily unique, since $\{\gamma_i\}$ can have degeneracy. If there exists an $r\times r$ matrix $D$ (generally not diagonal) that satisfies the following equation
\be
D = D^0+D^0 \tilde K D
\label{eq:D}
\ee
where $\tilde K=Q^\dagger K P$, then $\Gamma = P D Q^\dagger$ is the solution of the Bethe-Salpeter equation, since the solution to Eq. (\ref{eq:bse}), if it exists, is unique.   It is also implied that rank$(\Gamma) = r$, a fact not at all obvious at the outset.   Solving Eq. (\ref{eq:D}) by matrix inversion or iteration is much easier since $r\ll n$, provided the knowledge of $P$, $Q$ and $D^0$.

The direct singular value decomposition employing standard libraries~\cite{Golub65} requires an iterative procedure, and  it is estimated to have time  complexity $O(n^3)$ for a square matrix of order $n$.  Moreover, each iteration revisits a square matrix of order $n$ from the previous step, adding to the memory burden that can already be taxing, if not infeasible, for a large system. 
The fact that  $\Gamma^0$ has much lower rank than $n$ can again be exploited, which allows us to determine $P$ and $Q$ in a three-step process. A matrix $A\in \mathbb{C}^{n\times r}$ is generated randomly. Post-multiplication of $\Gamma^0$ by $A$ yields a projected matrix composed of $r$ column vectors in  ran($P$); and these vectors are subsequently orthogonalized by a matrix $R_1$ in a Gram-Schmidt process. Similarly, post-multiplication of ${\Gamma^0}^\dagger$ by $A$ yields a projected matrix composed of $r$ column vectors in ran($Q$), which is subsequently orthogonalized by a $R_2$. Both $R_1$ and $R_2$ are in $\mathbb{C}^{r\times r}$ and upper triangular. We have
\be
P_1 = PX;\;\; Q_1 = QY,
\label{eq:pq}
\ee
where $X=D^0Q^\dagger AR_1$, and $Y=D^0P^\dagger AR_2$, which are yet to be determined as $P$ and $Q$ are unknown. 

It is clear that $X$ and $Y$ are unitary, by the column orthogonality of $P$, $Q$ and $P_1$, $Q_1$. It follows from the unitarity of $X,Y$ and column orthogonality of $P,Q$ that
\be
P_1^\dagger \Gamma^0 Q_1 = X^\dagger D^0Y.
\ee
Evidently, this procedure reduces the original $n$-dimensional singular value decomposition to a task in $r$-dimensions. Subsequently, $P$ and $Q$ can be obtained via Eq. (\ref{eq:pq}). 
In fact, if $\Gamma^0$ and  $V$ are Hermitian, then $P=Q$ and we have the eigenvalue decomposition $\Gamma^0 = P D^0 P^{\dagger}$. The random matrix is projected only once by $\Gamma^0$ to get $P(Q)$,  which requires even less computational cost.
This method as outline above pivots on the projection from a complete linear space onto its subspaces, and we shall refer to it as the projective method for compact singular value decomposition.
As will be illustrated shortly, this projective method is practically stable and computationally efficient for our problem compared to the standard linear algebraic methods. 

\section{Stability and performance}
In this section we present basic sanity checks on the implementation of the above method for solving Eq. (\ref{eq:bse}) or Eq. (\ref{eq:bse01}). We will first show that  the projective compact singular value decomposition  has reasonable fidelity and stability in dealing with ill-conditioned problems. We will then show, with an example of 2-dimensional electron gas, that  our approach to the Bethe-Salpeter equation compares favorably with two of the direct methods in terms of how the computation time and memory scale with number of $\bm k$-points, $n_{k}$, in large $n_k$ limit. 

We now examine the stability of the projective algorithm for singular value decomposition described in the previous section. The stability is measured by the fidelity of the projective method in recovering the singular values for these matrices, which are ill-conditioned in degrees depending on the base $b$. Such tests are essential, especially in view of the relative instability of the Gram-Schmidt orthogonalization employed in this approach. In these numerical experiments, we randomly generate test matrices, which are then subject to singular value decomposition by the projective method as well as by standard singular value decomposition routines. The test matrices are all $\mathbb{C}^{3000\times 3000}$ with rank $r=25$, corresponding to 5-band systems, created in the following procedure: 

\noindent(1) $D^0$ is a diagonal matrix, whose diagonal elements are $\sigma_0^m=b^m, m=1,2,...,r$, and the values of the base $b= \tau,2,3,4$, where $\tau$ is the golden ratio;

\noindent(2) For each $D^0$ from step (1), a pair of orthogonal matrices, $P,Q\in \mathbb{C}^{n\times r}$. To generate $P$ or $Q$, a matrix in $\mathbb{C}^{n\times r}$ is generated at random with elements are picked from a uniform distribution inside the square on the complex plane with corners at $\pm(1+i)$, and then orthogonalized. The test matrix is then $PD^0Q^\dagger$.

\begin{figure}
	\includegraphics[width=62 mm]{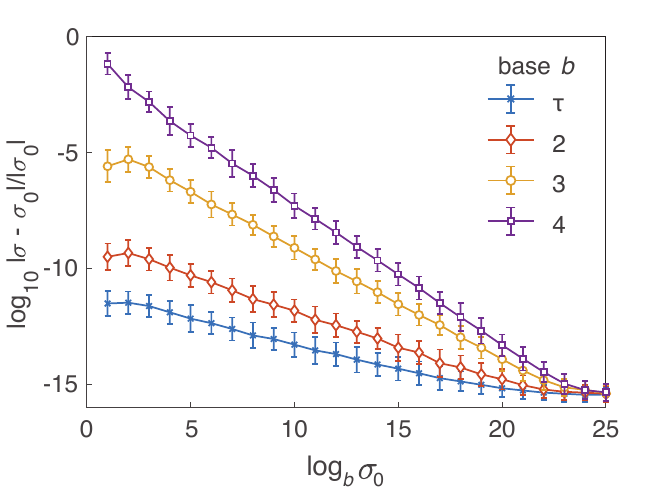}\centering
	\caption{\label{fig:002}Stability of the algorithm for ill-conditioned matrices. Here $\sigma^m_0 = b^m$ are the correct eigenvalues and $\sigma_m$ are the eigenvalues retrieved using the projective singular value decomposition described in the previous section. Vertical bars attached to each data points correspond the the standard deviation of the relative error.}
\end{figure}

The relative errors of the computed singular values is shown in Fig.~\ref{fig:002}. It is seen that for $b=\tau, 2$, the largest relative errors are less than $10^{-10}$, for exact singular values ranging over 7 orders of magnitude. Even in the case of $b=3$, where the singular values range over 12 orders of magnitude, the singular values retrieved has acceptable fidelity with relative errors less than  $10^{-5}$ . Only when the range of singular values is over $17$ orders of magnitude for $b=4$, the method becomes marginally unstable. These results indicate that this projective method should work as an expedient stratagem for the present application, although it may not offer sufficient stability to be a universally applicable approach. 

The computational efficiency of our approach to Bethe-Salpeter equation is analyzed here in comparison with other standard methods for linear equations, such as direct matrix inversion and biconjugate gradient method. Once $K$ and $\Gamma^0$ are prepared, the Bethe-Salpeter equation Eq. (\ref{eq:bse}) can be solved using the projective singular value decomposition method, in which matrix multiplications involving $\Gamma^0$ ($\Gamma^0A, P^{\dagger}\Gamma^0 P$) are the leading order operations with a time complexing of $O(rn^2)$. The matrix rank $r$ is dependent on the interaction range of the impurity potential as in Eq. (\ref{eq:r}). If localized impurity potentials with finite $n_r$ are applied to realistic materials with complex Fermi surfaces, we have $r< n_b^2n_r^2\ll n$ due to the large $n_k$  to sample the Brillouin zone. When $r\ll n$, this method reduces the time complexity from $O(n^3)$ to $O(n^2)$, a clear advantage not enjoyed by the other two methods. 
In terms of the random-access memory, the largest stored matrix ($P$) has the dimension $n\times r$, taking up less memory than $\Gamma^0$ that is required in the other two methods. Thus, as benefits from the dimensionality reduction of the matrices $A$ and $P$ for  small $r$, this algorithm is  desirable  in view of both time complexity and memory burden.

A simple example is now employed to gauge the efficiency of our method in comparison with the matrix inversion\footnote{Our code is compiled with the Intel(R) C Compiler XE 19.0 Update 1 for Linux*, and linear algebra routines from Intel Math Kernel Library are used unless otherwise noted. All the measurements were done on 2$*$Intel Xeon Gold 6130 with 2.10 GHz CPU, 16 core processors equipped with 256 GB physical RAM.} and an in-house  routine for the biconjugate gradient method. For this example, we use a one band system on a two-dimensional lattice with $H_{\bk}= \cos k_x+\cos k_y$. The impurity is introduced via Eq.~(\ref{eq:imp}). It is assumed here that the impurities are completely local, only permitting scattering within each lattice point with an amplitude $0<V<1$. The impurity concentration is  0.02, referring to the amount of impurities per lattice. The Bethe-Salpeter equation for the  maximally-crossed diagrams is solved for a total quasimomentum $\bm q=(0.008,0)$ at zero frequency. A uniform $\bm k$-point grid is adopted to sample the Brillouin zone. The disorder-averaged single-particle propagator is $G^A_{\bm k}= (\mu-H_{\bm k}-\Sigma^A)^{-1}$ with chemical $\mu=1$ and self energy $\Sigma^A=0.001\text i$. When using the matrix inversion method, the matrix $\Gamma^0$ is computed in full and stored on the random-access memory. In our projective singular value decomposition approach and the biconjugate gradient method, columns or blocks of $\Gamma^0$ split by quasimomentum are computed on-the-flight to reduce memory requirement. In our tests based on this single-band problem, a range of $n_k$ is used, which determines the dimension of $\Gamma^0$. $R_c$ values of 0, $2$ and $4$ are used, which lead to $\Gamma^0$ with different ranks.  

\begin{figure}
	\includegraphics[width=80 mm]{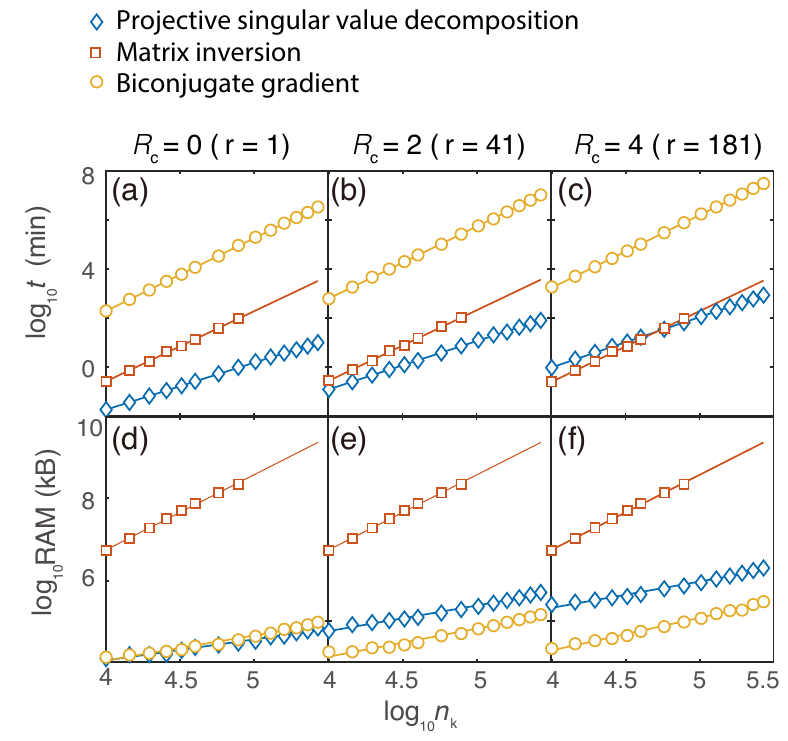}\centering
	\caption{\label{fig:time}The computational times and peak memories of the three algorithms versus {the number of $\bk$-points}, $n_k$.2-dimensional one-band Hamiltonian $H_{\bm k}=\cos k_x+\cos k_y$ is used. For the cutoff interaction range of impurity potential, three values are used: $R_c= 0,2,4$ (see Eq. (\ref{eq:imp})). For the rank of $\Gamma^0$, three values are used: $r=1,41,181$ respectively, all less than min\{$n_b^2n_k, n_b^2n_r^2$\} with  $n_r=1,13,49$. 
	}
\end{figure}

Panels (a)-(c) in Fig.~\ref{fig:time}  show the computation times as functions of $n_k$ on logarithmic scales, for $R_c= 0,2,4$ respectively. By counting the theoretical floating-point operations, we find that the computational time for our method is $O(n_k^2)$, and $O(n_k^3)$ for both matrix inversion and biconjugate-gradient method. The computational complexity is also confirmed empirically with a linear least squares of $\log_{10}t$ versus $\log_{10}n_k$.  While  all three methods yield identical results in all cases, the biconjugate gradient method is the most time consuming.  The estimated slopes of $\log_{10}t$-vs-$\log_{10}n_k$ for $R_c= 0,2,4$ using our method are about 2.0, whereas the slope is 2.9 for direct matrix inversion, and 3.0 for biconjugate gradient method. Thus, the empirical computational complexities agree well with theoretical expectations. Although direct matrix inversion method is advantageous for small $n_k$ large $R_c$, it ceases to be feasible for a moderate number of $\bm k$-points because of the overwhelming memory requirement for storing the matrices.

Fig.~\ref{fig:time} (d)-(f) show the peak memory load of the three methods. The amount of working memory required by matrix inversion increases rapidly with $n_k$, and ceases to be feasible very quickly for the compute node. Both biconjugate gradient and singular value decomposition methods require much less memory as $r\ll n_k$. The memory load of our method shows moderate increase with $R_c$ (or $r$). The above results  regarding the time and memory costs in Fig.~\ref{fig:time} show that the projective singular value decomposition method is both efficient  and memory-thrifty for solving the Bethe-Salpeter equation with a short-range impurity potential.

\section{Applications}
\begin{figure}[htbp]\centering 
	\includegraphics[width=80mm]{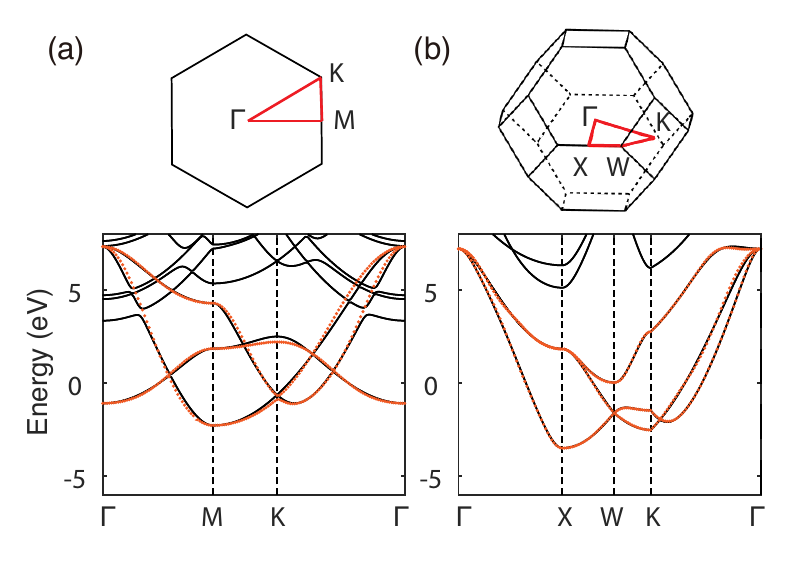}
	\caption{\label{fig:bandfit}Band structures of (a) monolayer Pb with a hexagonal lattice (b) bulk Pb. Energy dispersion from the tight-binding model (orange dots) is compared with the density-functional theory electronic structure (black lines).	} 
\end{figure}

For a  demonstration of the application to real materials, we use our method to evaluate the conductivity correction from maximally-crossed diagrams for 2-dimensional monolayer Pb with hexagonal lattice, and 3-dimensional fcc metal Pb. Here, we focus on the dc conductivity ($\omega=0$) at zero temperature.
Their tight-binding Hamiltonians are constructed with three $p$ orbitals of Pb using the Wannier90~\cite{Mostofi09} based on the electronic structure calculations based on density functional theory~\cite{Kresse1996}. As shown in Fig.~\ref{fig:bandfit}, the tight-binding models for monolayer Pb with hexagonal lattice, and bulk fcc Pb can perfectly capture the band dispersion around the Fermi level. A large number of $\bm k$-points are necessary in order to suitably sample the complicated Fermi surfaces of monolayer Pb with hexagonal lattice, and fcc Pb, which makes the direct inversion method infeasible for the Bethe-Salpeter equation, due to a huge memory requirement especially in 3-dimensional systems.

We start with the tight-binding Hamiltonian of Pb and the impurity scattering potential described by Eq. (\ref{eq:imp}). The disorder-averaged Green's functions $G_{\bm k}^{R/A}$ are obtained with an iterative method using  the self-consistent Born approximation(SCBA)~\cite{mattuck1974, Ando1998}, so that  the Ward identity~\cite{ward} is satisfied in the Bethe-Salpeter equation. In SCBA, the disorder averaged one-particle Green's function is given by 
\be \label{eq:G}
G_{\bk}^{R/A}=(\varepsilon-H_{\bk}-\Sigma_{\bk}^{R/A})^{-1},
\ee
in which the self energy owing the the presence of disorder is
\be\label{eq:selfe}
\Sigma_{\bk}^{R/A}=  \langle H'_{\bk\bk}\rangle  +\sum_{\bk'}\langle H'_{\bk\bk'}  G_{\bk'}^{R/A} H'_{\bk'\bk} \rangle.
\ee
In the SCBA, the chemical potential is self-consistently determined, in which the Brillouin sum to obtain the total electron number is facilitated by upsampling via Fourier interpolation of the poles of Green's functions.

With $G_{\bk}^{R/A}$ obtained from SCBA and the impurity potential, our projective singular value decomposition method can be used to solve the Bethe-Salpeter equations of maximally-crossed diagrams. The sum over these diagrams gives correction $\delta\sigma_{xx}= \sum_{\bm q}\delta\sigma_{xx}(\bm q)$, where
\be\label{eq:dsigma}
\delta\sigma_{xx}(\bm q)=\frac{e^2 \hbar}{2\pi \Omega} \sum_{\bk} v_{\bk}^xv_{\bq-\bk}^x G_{\bk}^RG_{\bk}^AG_{\bm q-\bk}^RG_{\bm q-\bk}^A\Gamma_{\bk,\bm q-\bk}(\bm q).
\ee
The orbital indices are omitted to highlight the momentum dependencies. The summation over $\bq$ is performed for $1/l_\phi<q<1/l$. Here,   $l$ is the mean-free path determined by the relaxation time $\tau$ and the diffusion coefficient $D$ which can be estimated from the calculations of Drude conductivity $l=\sqrt{D\tau}$. The coherence length, characterizing the inelastic scattering processes, $l_{\phi}$  is introduced as a parameter.

\begin{figure}
	\includegraphics[width=80mm]{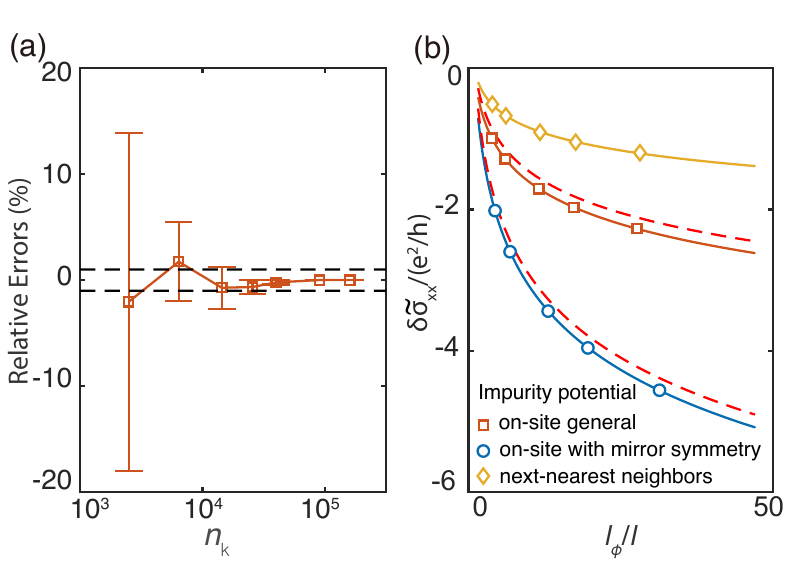}\centering
	\caption{\label{fig:lead2d}Computed vertex corrections to conductivity for 2-dimensional monolayer Pb with hexagonal lattice. (a) Relative errors of $\delta\sigma(\bm q)$ as a function of $n_k$ for a fully localized impurity potential. The vertical axis is the average of the relative error. The heights of vertical error bars correspond to twice the standard deviations. The black dashed lines labels $\pm 1\%$.  (b) The computed quantum correction to conductivity of Pb monolayer as a function of coherence length $l_{\phi}$ under three kinds of impurity potentials in form of Eq.~(\ref{eq:imp}): impurity potential with only on-site scattering, with on-site scattering respecting the mirror symmetry, and with next-nearest neighbor scattering. The corrections to conductivity are scaled to $\tilde{\sigma}_{xx}= \sqrt{\frac{{\sigma_{yy}}}{\sigma_{xx}}}\sigma_{xx}$ for comparison. 
		The open markers are the computed values and the solid curve is fitted from Eq.~(\ref{eq:lphi1}). 
		The conductivity correction directly from Eq.~(\ref{eq:lphi1}), and twice the value (see text),  are plotted as red dashed lines for reference.
	}
\end{figure}

For the 2-dimensional monolayer Pb with hexagonal lattice,  We first inspect the case of a fully-localized impurity potential described by Eq.~(\ref{eq:imp}) with a concentration of 0.0048, which induces only on-site hopping between $p$ orbitals with arbitrary numerical values shown in Table.~\ref{tab:1}. The $\bm q$-resolved correction $\delta\sigma(\bm q)$ is calculated to verify convergence with respect to $n_k$. Fig.~\ref{fig:lead2d}(a) shows the relative errors of $\delta\sigma(\bm q)$ versus $n_k$. The results of $\delta\sigma(\bm q)$ for the largest $n_k$($10^{5.2}$) are used as the estimate of correct values to calculate relative errors. It is seen that a large number of $\bk$-points up to $10^{4.6}$ are needed to achieve an accuracy within 1\%. Fig.~\ref{fig:lead2d}(b) displays the calculated vertex correction to conductivity as a function of  $l_{\phi}$. It is clear that $\delta\sigma_{xx}$ is negative, indicating weak localization. It can be fitted well by the anisotropic form of 2-dimensional weak localization~\cite{wolfle1984electron}
\be 
\delta\sigma_{xx}=-\frac{2e^2}{h\pi}\alpha \log\frac{l_{\phi}}{l},\label{eq:lphi1}
\ee
where $h$ is the Planck constant, the  factor of 2 results from the spin degeneracy, the coefficient $\alpha= \sqrt{\frac{{\sigma_{xx}}}{\sigma_{yy}}}$ contains the effects of anisotropy with $\sigma_{xx/yy}$ the Drude conductivity.  It implies that $\delta\sigma_{xx}(\bm{\tilde q})= -e^2/(4\pi^3h\tilde{q})^2$ if the momenta is rescaled  as $\bm{\tilde q} = (q_x, q_y/\alpha) $. Fitting the calculated $\delta\sigma_{xx}$ versus $l_{\phi}$ shown in Fig.~\ref{fig:lead2d}(b) using the formula in Eq.~(\ref{eq:lphi1}) gives $\alpha=1.02$ and $l=17.11 a_0$,  which are close to the $\sqrt{\frac{{\sigma_{xx}}}{\sigma_{yy}}}=1.01$  and the given mean-free path  $l=20.66a_0$. The small mismatch~\cite{Wei2018} between the fitted and the theoretically derived values of $\alpha$ and $l$ originates from the deviation of the calculated $\delta\sigma_{xx}(\bm{\tilde q})$  with the result based on second-order perturbation theory~\cite{wolfle1984electron}, $-e^2/(4\pi^3h\tilde{q})^2$.

\begin{table}\centering
	\caption{Impurity-induced hopping matrix elements in eV.}
	\setlength{\tabcolsep}{7mm}{
		\begin{tabular}{cccc}
			\hline\hline
			Orbital & $p_x$  & $p_y$  & $p_z$  \\ \hline
			$p_x$       & 0.5 & 2.8 & 0.8 \\ 
			$p_y$       & 2.8 & 0.5 & 0.8 \\
			$p_z$       & 0.8 & 0.8 & -1  \\ 
			\hline\hline
	\end{tabular}}\label{tab:1}
\end{table}

We  now turn our attention to two more types of impurity:  one with a finite interaction range, and the other with higher symmetry.  We find that in these cases, the theory summarized in Eq. (\ref{eq:lphi1}) fails.
In the former case, the hopping range is extended to the  next-nearest neighbors from the localized potential $V_{ab}(0,0)$ in Table.~\ref{tab:1} by $V_{ab}{(\bm R,\bm R')}=V_{ab}(0,0)\sqrt[3]{ \frac{\rm{sin} R}{R}\frac{\rm{sin} R'}{R'}\frac{{\rm{sin}} |\bm R-\bm R'|}{|\bm R-\bm R'|}}$.  Similarly, the results of $\delta\sigma_{xx}$ shown in Fig.~\ref{fig:lead2d}(b) can also be described by Eq.~(\ref{eq:lphi1}) with $\alpha= 0.65$ and $l = 18.01a_0$. Notably, however, $\alpha$  differs significantly from the theoretical value $\sqrt{\frac{{\sigma_{xx}}}{\sigma_{yy}}}=1.19$. The discrepancy  indicates that the effectiveness of the second-order perturbation is seriously  compromised when the impurity potential is not fully localized and has a variation in the $\bk$-space. In the final case, the impurity induced hopping is again fully local, but it is assumed that impurity potential respects the mirror symmetry of the plane of the hcp lattice. In this case, the hopping matrix elements are the same with Table.~\ref{tab:1}, except that the hopping between $p_z$ and ($p_x,p_y$) orbitals is forbidden. The fitted  $l = 16.47a_0$ is close to the given value $18.23a_0$, while the fitted $\alpha=2.00$  is twice the $ \sqrt{\frac{{\sigma_{xx}}}{\sigma_{yy}}}=0.99$ as shown in Fig.~\ref{fig:lead2d}. This deviation from  Eq.~(\ref{eq:lphi1})  corresponds to adding two decoupled parts derived from $p_z$ and ($p_x,p_y$) orbitals, respectively, and hence the simple doubling of $\alpha$. Therefore, the nature of impurity potential, in particular, its range, orbital and  symmetry, plays a crucial role in quantum correction to conductivity. They could render the simple theories like Eq. (\ref{eq:lphi1}) ineffective. Careful numerical approach becomes necessary in the investigation of quantum correction in realistic materials with complicated impurity scattering.  

\begin{figure}[htp]\centering
	\includegraphics[width=45mm]{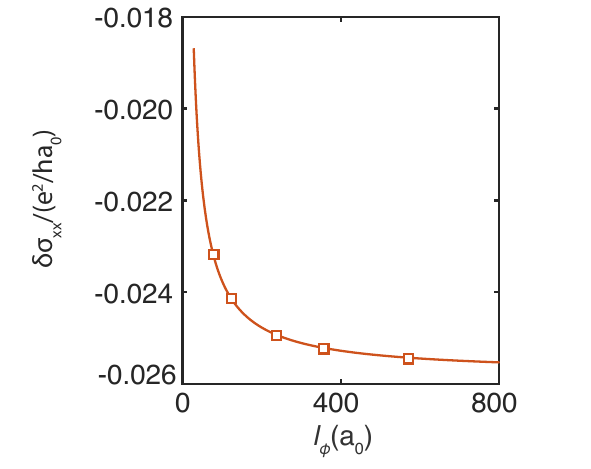}
	\caption{\label{fig:lead3d}Computed vertex corrections for fcc Pb as a function of coherent length $l_{\phi}$. Here $l_\phi$ is given in the units of the shortest Pb-Pb contact. }
\end{figure}

The quantum correction to conductivity of fcc Pb is shown in Fig.~\ref{fig:lead3d} with $R_c=0$, $V_{ab}=5.0\delta_{ab}$  eV, and $n_i/n_k=0.02$. The computed weak localization correction to conductivity can be fitted well by the equation for a 3-dimensional metallic system~\cite{BERGMANN19841}
\be
\delta\sigma_{xx}=\frac{2e^2}{h\pi^2}(1/l_{\phi}-1/l),\label{eq:lphi}
\ee
Fitting the calculated $\delta\sigma_{xx}$ versus $l_{\phi}$ shown in Fig.~\ref{fig:lead3d} to formula Eq.~(\ref{eq:lphi}) gives the fitted $l=7.86a_0$. It is slightly larger than the given mean-free path  $11.01a_0$, which is likely to reflect higher order effects not accounted for by the second-order perturbation theory behind  Eq. (\ref{eq:lphi})~\cite{wolfle1984electron}.	
These two examples illustrate that the non-iterative approach to the Bethe-Salpeter equation can be successfully applied to the calculation of vertex correction for multi-band Hamiltonian derived from realistic materials.

\section{Discussions and summary}
We have presented a non-iterative method for solving the Bethe-Salpeter equation arising in the vertex correction for electric conductivity. This method exploits the fact when the scattering potential is short-ranged, the scattering vertex is typically not rank full, allowing for reduction of matrix dimensions through an expedient projective singular value decomposition. We have shown that this projective singular value decomposition offers sufficient stability and robustness for this problem. The resultant algorithm for the Bethe-Salpeter equation has a theoretical computational complexity of \ $O(n_k^2)$ when the rank of the bare scattering vertex is much smaller than the number of $\bm k$-points required for sampling the Fermi surface. It is empirically established, by investigating the case of single band model in 2 dimensions with short-ranged impurity potentials, that our method compares favorably to conventional linear algebraic approaches with a computational complexity of $O(n_k^3)$, namely, matrix inversion and biconjugate gradient method. 

It is then demonstrated that the method can be successfully applied to compute the quantum correction for 2-dimensional and 3-dimensional systems, for which the tight-binding Hamiltonians are obtained from DFT calculations with ab initio accuracy.  Our method allows the introduction of various types of impurity potentials. It is found that the calculated quantum correction to conductivity departs qualitatively from the expectation of a second-order perturbation theory. This correction is signally modulated by the the range, orbital and symmetry of the impurity potential, indicating that the nature of impurity is essential and requires careful considerations in real materials.

Thus, our method provides an efficient machinery for evaluating the quantum effects in conductivity of materials, based on realistic electronic structures obtained from accurate electronic structure methods. We expect that this newly proposed method could have a profound impact in studying transport properties by Kubo formula. Nonetheless, it may be emphasized that at this point, not all the inputs into our method are obtained ab initio. First, the Hamiltonian describing how the electrons scatter with an impurity has to be given as a model at this stage. Methods should be developed to evaluate the impurity scattering potentials~\cite{Talwar1982,Kemper2009,Nakamura2011}.  Second, the coherence length needed for the momentum cutoff is also entered as a given parameter~\cite{BERGMANN19841}.

\section*{Acknowledgments}
This work was supported by the National Natural Science Foundation of China (Grants No. 11725415 and No. 11934001), the Ministry of Science and Technology of the People’s Republic of China (Grants No. 2018YFA0305601 and No. 2016YFA0301004), and by Strategic Priority Research Program of Chinese Academy of Sciences, Grant No. XDB28000000.
\bibliographystyle{statto}
\bibliography{bse}

\end{spacing}
\end{document}